\documentclass[11pt]{article}
\pdfoutput=1
\usepackage{jcappub,natbib,float,verbatim}
\usepackage{color}
\usepackage{comment,subcaption}
\usepackage{tikz,booktabs}
\newcommand{\head}[1]{\textnormal{\textbf{#1}}}
\title{Measuring angular diameter distances of strong gravitational lenses}
\author{I. Jee$^{a}$, E. Komatsu$^{a,b}$, S. H. Suyu$^{c}$}
\affiliation[a]{Max-Planck-Institut f\"ur Astrophysik, Karl-Schwarzschild Str. 1, 85741 Garching, Germany}
\affiliation[b]{Kavli Institute for the Physics and Mathematics of the
Universe, Todai Institutes for Advanced Study, the University of Tokyo,
Kashiwa, Japan 277-8583 (Kavli IPMU, WPI)}
\affiliation[c]{Institute of Astronomy and Astrophysics, Academia Sinica,
P.O. Box 23-141, Taipei 10617, Taiwan} 
\emailAdd{ijee@mpa-garching.mpg.de}
\abstract{%
The distance-redshift relation plays a fundamental role in constraining
cosmological models. In this paper, we show that measurements of
positions and time delays of strongly lensed images of a background
galaxy, as well as those of the velocity dispersion and mass profile
of a lens galaxy, can be combined to extract the angular diameter
distance of the lens galaxy. Physically, as the velocity
dispersion and the time delay give a gravitational potential ($GM/r$)
and a mass ($GM$) of the lens, respectively, dividing them gives a
physical size ($r$) of the lens. Comparing the physical size with the
image positions of a lensed galaxy gives the angular diameter distance
to the lens. A mismatch between the exact locations at which these
measurements are made can be corrected by measuring a local slope of the
mass profile. We expand on the original idea put forward by Paraficz
and Hjorth, who analyzed singular isothermal lenses, by allowing for an
arbitrary slope of a power-law spherical mass density profile, an
external convergence, and an anisotropic velocity dispersion. We find
that the effect of external convergence cancels out when dividing the time delays 
and velocity dispersion measurements. We
derive a formula for the uncertainty in the angular diameter distance in
terms of the uncertainties in the observables. As an application, we
use two existing strong lens systems, B1608+656 ($z_{\rm L}=0.6304$) and
RXJ1131$-$1231 ($z_{\rm L}=0.295$), to show that the uncertainty in the inferred angular
diameter distances is dominated by that in the velocity
dispersion, $\sigma^2$, and its anisotropy. We find that the current
data on these systems should yield about 16\% uncertainty in
$D_A$ {\it per object}. This improves to 13\% when we measure
 $\sigma^2$ at the so-called sweet-spot radius. Achieving 7\% is
possible if we can determine $\sigma^2$ with 5\% precision.
}
\begin{document}
\maketitle
\flushbottom
\section{Introduction}
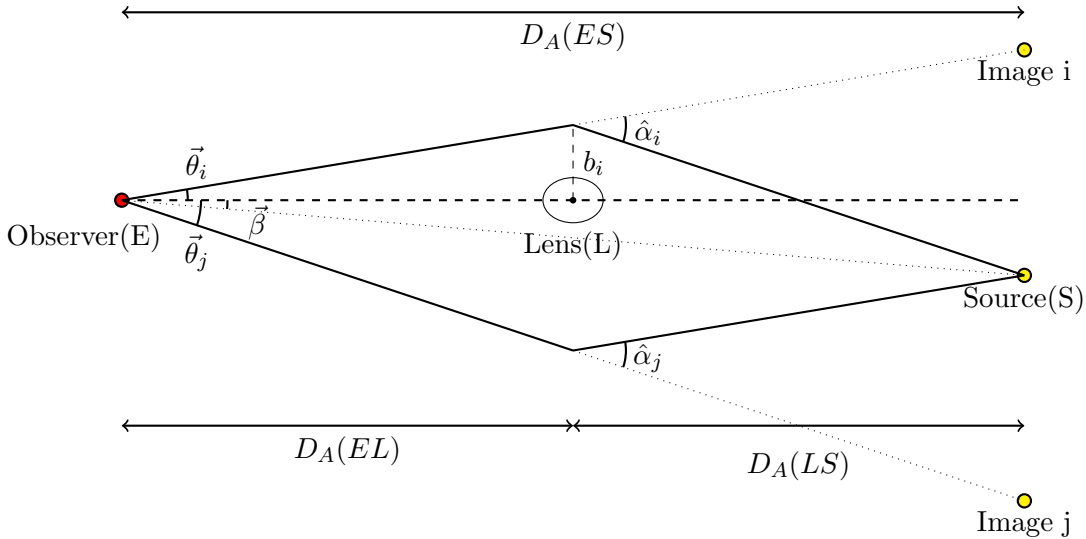
\begin{figure}[t]
\begin{center}
\begin{tikzpicture}
\tikzstyle{vertex}=[fill,thick,circle,minimum size=2.5pt,inner sep=0pt]
\tikzstyle{object}=[fill=yellow,thick,draw,circle,minimum size=5pt,inner sep=0pt]
\draw(12,4) node[object](u){};
\draw(12,-2) node[object](v){};
\draw(12,1) node[object](w){};
\draw(0,2) node[fill=red,thick,draw,circle,minimum size=5pt, inner sep=0pt](x){};
\draw(-0.5,1.5) node{Observer(E)};
\draw(6,1.4) node{Lens(L)};
\draw(12,1) node[anchor=north]{Source(S)};
\draw(12,4) node[anchor=north]{Image i};
\draw(12,-2) node[anchor=north]{Image j};
\draw(1,2.5) node{$\vec{\theta}_i$};
\draw(1,1.3) node{$\vec{\theta}_j$};
\draw(1.8,1.7) node{$\vec{\beta}$};
\draw(7,2.9) node{$\hat{\alpha}_i$};
\draw(7,-0.1) node{$\hat{\alpha}_j$};
\draw(3,-1)  node[anchor=north]{$D_A(EL)$};
\draw(9,-1.2) node[anchor=north]{$D_A(LS)$};
\draw(6,4.5) node[anchor=north]{$D_A(ES)$};
\draw(6,2.5) node[anchor=west]{$b_i$};
\draw[thick,dashed](0,2)--(12,2);
\draw(6,2) ellipse (0.4cm and 0.3cm);
\draw(6,2) node[vertex](u) {};
\draw[thick](0,2)--(6,3);
\draw[thick](6,3)--(12,1);
\draw[dotted](6,3)--(12,4);
\draw[thick](0,2)--(6,0);
\draw[thick](6,0)--(12,1);
\draw[dotted](6,0)--(12,-2);
\draw[dotted](0,2)--(12,1);
\draw[thick,<->](0,-1)--(6,-1);
\draw[thick,<->](6,-1)--(12,-1);
\draw[thick,<->](0,4.5)--(12,4.5);
\draw[dashed](6,2)--(6,3);
\begin{scope}
\path[clip] (6,3) -- (0,2) -- (6,2) -- cycle;
\node[circle,draw,minimum size=50pt,thick] at (0,2) (circ) {};
\end{scope}
\begin{scope}
\path[clip] (6,2) -- (0,2) -- (6,0) -- cycle;
\node[circle,draw,minimum size=60pt,thick] at (0,2) (circ) {};
\end{scope}
\begin{scope}
\path[clip] (6,2) -- (0,2) -- (12,1) -- cycle;
\node[circle,draw,minimum size=80pt,thick] at (0,2) (circ) {};
\end{scope}
\begin{scope}
\path[clip] (12,-2) -- (6,0) -- (12,1) -- cycle;
\node[circle,draw,minimum size=40pt,thick] at (6,0) (circ) {};
\end{scope}
\begin{scope}
\path[clip] (12,4) -- (6,3) -- (12,1) -- cycle;
\node[circle,draw,minimum size=40pt,thick] at (6,3) (circ) {};
\end{scope}
\end{tikzpicture}
\caption{Configuration of a strong lens system, with definition of the
 variables used throughout this paper. All angles are measured with
 respect to the center of the lens galaxy; $\vec{\theta}$ is the angular
 position of \textbf{}the image; $\vec{\beta}$ is the angular position
 of the source in the absence of the lens; $\vec{\alpha}$ is the scaled
 deflection angle; $\hat{\alpha}$ is the deflection angle at the lens
 plane; and $\vec{b}$ is the physical separation to the closest approach
 at the lens plane.}
\label{fig:lens_config}
\end{center}
\end{figure}

Individual strong gravitational lens systems can be used to measure
cosmological parameters via a combination of the cosmological distances
\cite{im/griffiths/ratnatunga:1997,Futamase/hamana:1999,Futamase/yoshida:2001,Yamamoto/futamase:2001,linder:2004}. Recently,
a particular combination of the distances called the ``time-delay distance''
of strongly lensed time delay systems has yielded precise
determinations of the Hubble constant
\cite{koopmans/etal:2003,suyu/others:2010,suyu/others:2012}. The
time-delay distance is the angular diameter distance to the lens from Earth, $D_A(EL)$, multiplied
by the distance to the source, $D_A(ES)$, divided by the distance
between the lens and the source, $D_A(LS)$. While this
combination is sensitive to the Hubble constant, it is less so
to the other cosmological parameters than $D_A(EL)$ itself
\cite{Fukugita/kasai:1990}. See figure \ref{fig:lens_config} for the
definition of these distances. 

To extract more cosmological information acquirable from
the strong lens time delay systems, Paraficz and Hjorth
\cite{paraficz/hjorth:2009} have shown that, by assuming the density profile
of the lens galaxy, one can obtain $D_A(EL)$ from time-delay lenses.
The basic physics behind this idea is simple: the velocity
dispersion gives the depth of the potential at the point where it is measured,
and the time delay gives the mass of the lens galaxy enclosed within the
position at which images are formed. Thus, dividing them gives the
physical size of 
the system. We can then estimate $D_A(EL)$ by dividing the physical size by
the angular separation of lensed image positions. Their
analysis was limited to the singular isothermal sphere (SIS) density
profile, as well as to an isotropic velocity dispersion. In this paper,
we show that this simple physical picture holds even when
we extend the analysis by including an arbitrary power-law profile, the
effect of external convergence, and an anisotropic velocity
structure. We show explicitly how to extract $D_A(EL)$ from the
observational data, and provide an estimate of its associated uncertainty.

The rest of the paper is organized as follows. In
section~\ref{sec:basic}, we present the basic idea using a simplified
SIS model, following ref.~\cite{paraficz/hjorth:2009}. In
section~\ref{sec:ouranalysis}, we expand on
ref.~\cite{paraficz/hjorth:2009} by allowing for an arbitrary slope of a
power-law spherical  mass density profile and external convergence. 
In section~\ref{sec:implications}, we derive an analytical
formula relating the uncertainty in $D_A(EL)$ to the uncertainties in
the observable quantities, and apply the formula to the observed strong
lens time delay systems, B1608+656 and RXJ1131$-$1231. In
section~\ref{sec:anisotropy}, we use
Monte-Carlo simulations to study the effect of anisotropic velocity
dispersion on the uncertainty in $D_A(EL)$. We conclude in
section \ref{sec:discussion}. In the appendix, we show how General
Relativity allows us to calculate the deflection angle at the lens
plane.

\section{Basics of the analysis} 
\label{sec:basic}
 \subsection{The idea: a simple analysis using singular isothermal
 spheres}
\label{sec:sis}
We review the basic idea with the simplest case in which the mass
density profile of a lens galaxy is given by an SIS. This case has been
worked out by Paraficz and Hjorth in 2009
\cite{paraficz/hjorth:2009}. The density distribution of an SIS lens,
$\rho_\mathrm{SIS}$, is given by
\begin{equation}
 \rho_\mathrm{SIS}(r)=\frac{\sigma^2}{2\pi G r^2},
 \label{eq:rho_sis}
\end{equation}
where $\sigma^2$ is the three-dimensional isotropic velocity
dispersion. The Einstein ring radius, $\theta_E$, is related to
$\sigma^2$ via
\begin{equation}
\sigma^2=\theta_\mathrm{E}  \frac{c^2}{4\pi} \frac{D_A(ES)}{D_A(LS)}.
\label{eq:sig}
\end{equation}
Clearly, the relation between the two observable quantities, $\theta_\mathrm{E}$
and $\sigma$, depends on the distance {\it ratio}.

To extract the actual angular diameter distance to the lens, $D_A(EL)$,
instead of the ratio, we need to include the lensing time delay
\citep{refsdal:1964}. The presence of intervening mass between the
observer and the source, usually galaxies and/or clusters of galaxies,
causes two different components on time delay: the {\it geometrical time
delay} and the {\it potential time delay}. Strongly lensed systems show
multiple images as photons coming from the source take different paths:
images are located at the closest approach to the lens of each path. The
geometrical part of the time delay is caused by the fact that the total
path lengths differ, while the potential part is caused by the
difference in the depths of potential at each image position of the
path. 

In a SIS lens, the time delay between
two images can be written as  
\begin{equation}
\Delta t_{i,j}\equiv t_i-t_j=
\frac{1+z_\mathrm{L}}{2c}\frac{D_A(EL) D_A(ES)}{D_A(LS)} (\theta_{j}^2-\theta_{i}^2),
\label{eq:SISdt}
\end{equation}
where $\theta_{i}$ is the angular separation between the $i$-th
image and the center of the lens galaxy, and $t_i$ is the
absolute time delay of the $i$-th image, i.e., the delay in comparison to 
the case where the lens is absent \cite{witt/mao/keeton:2000}. The distance ratio
that appears in this relation is the time-delay distance, $D_{\Delta
t}\equiv (1+z_\mathrm{L}){D_A(EL) D_A(ES)}/{D_A(LS)}$, which depends
primarily on $H_0$ and has a limited sensitivity to the other
cosmological parameters, such as the equation of state of dark energy.

Remarkably, when we combine the above equation with equation
(\ref{eq:sig}) and $\theta_E=(\theta_{i}+\theta_{j})/2$, we
obtain the angular diameter distance to the lens:
\begin{equation}
D_A(EL) (\theta_{j}-\theta_{i} )= \frac{c^3\Delta
t_{i,j}}{4\pi\sigma^2(1+z_\mathrm{L})}. 
\label{eq:D_A_SIS}
\end{equation}
The physical interpretation of the above analysis is as follows: the
velocity dispersion is determined by the gravitational potential of the
lens, $GM/r$. The time delay gives the mass of the lens system, $GM$,
and thus dividing them gives the physical size of the system, $r$. Since
the angular scale of the system is directly observable via
$\theta_j-\theta_i$, one can estimate the angular diameter distance to
the lens. Equation~(\ref{eq:D_A_SIS}) indeed gives the angular
diameter distance as $D_A(EL)\propto
\Delta t_{i,j}/[\sigma^2(\theta_{j}-\theta_{i})]$;
thus, the uncertainty in $D_A(EL)$ is given by the quadrature sum of
the uncertainties in the time delay, velocity dispersion, and image
position measurements. 

As the velocity dispersion uncertainty is
usually the biggest of all uncertainties,  the uncertainty in $D_A(EL)$
is expected to be dominated by the velocity dispersion uncertainty.
The goal of this paper is to extend this analysis to more general lenses.

\subsection{Lensing theory and equations}
\label{sec:basic_lensing}
Before we proceed, let us review some of  general equations for
strong lensing, following ref.~\cite{suyu/others:2010}.
Let the angular position of the image be $\vec{\theta}$ and that of
the source be $\vec{\beta}$, as shown in fig.~\ref{fig:lens_config}. The absolute time delay can be written as 
\begin{equation}
t(\vec{\theta},\vec{\beta}) = \frac{1}{c}(1+z_\mathrm{L})\frac{D_A(EL)D_A(ES)}{D_A(LS)}\phi(\vec{\theta},\vec{\beta}),
\label{eq:td}
\end{equation}
where $\phi$ is the so-called Fermat potential, which is defined as
\begin{equation}
\phi(\vec{\theta},\vec{\beta}) \equiv \frac{(\vec{\theta}-\vec{\beta})^2}{2}-\psi(\vec{\theta}).
\label{eq:pot}
\end{equation}
The first and the second terms in equation~(\ref{eq:td}) are geometrical
and potential time-delay terms, respectively. Here, $\psi$ is the lens
potential, which is calculated as
\begin{equation}
\psi(\vec{\theta})  = \frac{1}{\pi} \int d^2\theta' \kappa(\vec{\theta}')\ln |\vec{\theta}-\vec{\theta}'|,
\end{equation} 
where the lensing convergence field, $\kappa$, is defined by
\begin{equation}
\kappa(\vec{\theta}) \equiv
 \frac{\Sigma(\vec{\theta})}{\Sigma_\mathrm{cr}}. 
\label{eq:convergence-def}
\end{equation}
The projected surface mass density, $\Sigma$, is
\begin{equation}
\Sigma(\vec{\theta}) = \int_{-\infty}^{\infty} \rho[D_A(EL)\vec{\theta},\ell] ~d\ell,
\label{eq:kappa_rho}
\end{equation}
where $\ell$ denotes the line-of-sight coordinate, and 
\begin{equation}
\Sigma_\mathrm{cr}\equiv \frac{c^2 }{4 \pi G }\frac{D_A(ES)}{D_A(EL) D_A(LS)},
\end{equation}
is the critical surface mass density. Physically, when $\kappa>1$, the system
satisfies the sufficient condition to form multiple images.

The absolute time delay, $t$, is not an observable as we
cannot directly observe the source without the lens, or the time
difference between lensed and un-lensed images. However, if we have
multiple images, we can compare the
relative time delay between image pairs to calculate the time delay
between two (or more) lensed images. Also, $\phi$ can be modeled to
satisfy observational constraints such as image positions, flux
ratios and time-delay differences between multiple pairs of images; thus,
we can obtain the time-delay distance.

In a differential form, the lens potential is related to the convergence
field via
\begin{equation}
\kappa(\vec{\theta})=\frac{1}{2} \nabla ^2 \psi(\vec{\theta}),
\end{equation}
where $\nabla$ is a derivative in $\vec{\theta}$ coordinates.
Now we can write the lens equation which relates the observed image position to the source position
in terms of the lens potential,
\begin{equation}
\vec{\theta}-\vec{\beta} = \nabla \psi(\vec{\theta})=\vec{\alpha},
\label{eq:lens}
\end{equation}
where $\vec{\alpha}$ is the scaled deflection angle.

\section{More realistic lenses}
\label{sec:ouranalysis}
The analysis in section \ref{sec:sis} assumes the simplest possible lens system:
an SIS density profile with an isotropic velocity
dispersion. While the SIS profile is widely used to model lens galaxies and is
considered  
as a good approximation, several studies have shown that 
slopes of density profiles of individual galaxies show a non-negligible
scatter from the SIS \cite{koopmans/etal:2009,auger/etal:2010,barnabe/etal:2011, sonnenfeld/others:2013}.
In this section, we consider an arbitrary power-law density
profile (section~\ref{sec:power-law}) to show that, in such a model, we can still extract $D_A(EL)$
from $\Delta t_{i,j}$, $\sigma^2$, and image positions. We then show
that the external convergence cancels out (section~\ref{sec:ext_conv}). We note that 
spherical symmetry is assumed throughout the paper.

\subsection{Arbitrary slope of the spherical lens mass profile}
\label{sec:power-law}
Studies of early type galaxies (ETGs) as lenses have shown that the averaged \textit{total} mass density profiles can be well approximated as a power-law, and also typical ellipticity of galaxies is fairly small \cite{koopmans/etal:2009, auger/etal:2010, barnabe/etal:2011, sonnenfeld/others:2013,cappellari/etal:2015}. Thus we allow the total mass density of a lens to follow a general power-law with spherical symmetry:
\begin{equation}
\rho = \rho_0 \Big(\frac{r}{r_0}\Big)^{-\gamma'}.
\label{eq:density}
\end{equation}
The distribution becomes a SIS for $\gamma' = 2$ (section \ref{sec:sis}).
The lens potential also has a power-law form, $\psi(\theta)
\propto \theta^l$, with  $l= 3-\gamma'$. The scaled deflection, $\vec{\alpha}$, which is given by $\nabla \psi=\vec{\alpha}$, and the lens equation, 
$\vec{\beta} = \vec{\theta}-\vec{\alpha}$, gives
\begin{equation}
\psi = \frac{1}{l}~\vec{\theta}\cdot (\vec{\theta} - \vec{\beta}).
\end{equation}
Using this result  in  equation (\ref{eq:td}), we obtain the time delay
between two images as
\begin{equation}
\Delta t_{i,j} = \frac{1+z_{\rm L}}{2c}\frac{D_A(EL) D_A(ES)}{D_A(LS)}
\left\{(\vec{\theta_{i}}-\vec{\beta})^2 - (\vec{\theta_{j}}-\vec{\beta})^2 - \frac{2}{l}\left[\vec{\theta_{i}}\cdot(\vec{\theta_{i}}-\vec{\beta})-\vec{\theta_{j}}\cdot(\vec{\theta_{j}}-\vec{\beta})\right]\right\}.
\label{eq:dt_new}
\end{equation}
From the geometry of the system, the lens equation and the definition of the angular diameter distance, 
the following relation between 
$\vec{\theta}$, $\vec{\beta}$, and $\hat{\alpha}$ holds:
\begin{equation}
\vec{\theta}-\vec{\beta}=\vec{\alpha}=\frac{ D_A(LS)}{D_A(ES)} \hat{\alpha},
\end{equation}
where $\hat{\alpha}$ is the deflection angle at the lens plane. 
We substitute $\vec{\theta}-\vec{\beta}$ in equation (\ref{eq:dt_new}) for $\hat{\alpha}$, and write 
\begin{equation}
\Delta t_{i,j} = D_A(EL)\frac{(1+z_\mathrm{L})}{2c}\left[ (\hat{\alpha}_{i} + \hat{\alpha}_{j})\cdot(\vec{\theta}_{i} - \vec{\theta}_{j}) -\frac{2}{l}(\vec{\theta_{i}}\cdot\hat{\alpha}_{i} -\vec{\theta_{j}}\cdot \hat{\alpha}_{j} )\right].
\label{eq:dt_delta}
\end{equation}
The remaining task is to relate $\hat{\alpha}$ to observables. As the potential of a spherically symmetric system only has a radial component with respect to the center, $\vec{\alpha}$,
$\vec{\beta}$ and $\vec{\theta}$ have only radial components. 
Let us define $\alpha \equiv |\hat{\alpha}|$, which is the magnitude of the deflection angle at the lens plane.
Under the power-law density profile model, $\alpha$ is given by
\begin{equation}
\alpha(b)  = \frac{2GM(b)}{c^2b}F(\gamma')\propto b^{-\gamma'+2},
\end{equation}
where $b$ is the physical separation between the lens and
the point of the closest approach of the light ray, and
\begin{equation}
F(\gamma') \equiv \frac{\sqrt{\pi}\Gamma\left[\frac{1}{2}(-1+\gamma')\right]}{\Gamma(\frac{\gamma'}{2})}.
\end{equation}
The derivation of this formula is given in appendix \ref{sec:def_angle}.

Using the virial theorem, we obtain the radial velocity
dispersion at a given radius $r$ as
\begin{equation}
\sigma_r^2(r) = \frac{1}{2(\gamma' -1)} \frac{GM(r)}{r}\propto r^{-\gamma'+2}.
\label{eq:sig_r}
\end{equation}
If the velocity dispersion is isotropic,
$\sigma_r^2(r)=\frac13\sigma^2(r)$, and the radial velocity
dispersion is the same as the line-of-sight velocity dispersion, which
is observable. As both $\alpha$ and $\sigma_r^2(r)$ scale with radii in
the same way, we can write $\alpha(b)$ as
\begin{equation}
\alpha = \frac{4(\gamma'-1)}{c^2} F(\gamma')\sigma^2_r(b) = \frac{4(\gamma'-1)}{c^2}
 F(\gamma')\sigma^2_r(r)\left(\frac{b}{r}\right)^{-\gamma'+2}.
\label{eq:alpha}
\end{equation}
We then obtain $D_A(EL)$ from equation (\ref{eq:dt_delta}) with $\alpha$ given by equation (\ref{eq:alpha}), 
\begin{equation}
D_A(EL) = \frac{c^3\Delta
t_{i,j}}{4\pi\sigma_r^2(r)(1+z_{\rm{L}})} (\Delta\tilde{\theta}_{i,j})^{-1},
\label{eq:power-law}
\end{equation}
where\footnote{We use $\hat{\theta_i}\cdot\hat{\theta_j}=-1$ in reducing
the vector dot products in equation (\ref{eq:dt_delta}) to the scalar products
in equation (\ref{eq:Delta_theta_ij}).}
\begin{equation}
 (\Delta\tilde{\theta}_{i,j})^{-1}\equiv
  \frac{2\pi\left\{\frac{2}{-\gamma'+3}\left[
					{\theta_{j}}\left(\frac{\theta_{j}}{\Theta}\right)^{-\gamma'+2}
-{\theta_{i}}\left(\frac{\theta_{i}}{\Theta}\right)^{-\gamma'+2}
\right]+
(\theta_{i} + \theta_{j})\left[\left(\frac{\theta_{i}}{\Theta}\right)^{-\gamma'+2}-\left(\frac{\theta_{j}}{\Theta}\right)^{-\gamma'+2}\right]
\right\}^{-1}}{F(\gamma') (\gamma'-1)},
\label{eq:Delta_theta_ij}
\end{equation}
and $\Theta$ is the angular position at which 
the velocity dispersion is measured, i.e., $r = \Theta D_A(EL)$.
For $\gamma' = 2$, we obtain
$\Delta\tilde{\theta}_{i,j}=\theta_j-\theta_i$, and thus we can
reproduce the result of the SIS model (equation (\ref{eq:D_A_SIS})).

Equation~(\ref{eq:power-law}) still supports the basic physical picture
that the ratio of $\Delta t_{i,j}$ and $\sigma^2_r$ gives some effective
physical size of the lens, and dividing it by the appropriate angular
separation in the sky, $\Delta\tilde{\theta}_{i,j}$, gives the angular
diameter distance. The main difference between the SIS and the power law
density profiles is that, in the latter case, the velocity dispersion is
a function of radii. In general, image positions are different from the
points at which the velocity dispersion is measured. Thus, we need to
 correct for the mismatch of the exact
locations of the velocity dispersion measurement and the image
positions. This is why the
$\left(\frac{\theta}{\Theta}\right)^{-\gamma'+2}$ term appears in the
final expression of $D_A(EL)$: it scales the velocity dispersion such
that we can get the potential at the image
position. This requires us to measure (or model) the density slope,
$\gamma'$, as well.

\subsection{External convergence}
\label{sec:ext_conv}
In modeling realistic lens systems, one important factor to consider is
the so-called ``mass-sheet transformation (MST)''. MST is a subset of
the source-position transformation \cite{schneider/sluse:2013}. Degeneracy exists, such that there
are many mass models of the lens galaxy that can simultaneously
reproduce most of the lensing observables, such as image positions and
flux ratios, with different source positions
\cite{falco/others:1985}. This degeneracy constitutes one of the dominant 
sources of uncertainty in measuring the time-delay distance 
\cite{suyu/others:2010,suyu/others:2012,suyu/others:2013,
schneider/sluse:2013}. 
In this subsection, we show that
the effect of MST cancels out, leaving no effect on the
inferred $D_A(EL)$. 

Once we choose a model for the convergence field,
$\kappa_\mathrm{model}(\vec{\theta})$, that matches the observations, we
transform $\kappa_\mathrm{model}$ and $\vec{\alpha}$ to obtain 
a new convergence field, $\kappa_\mathrm{MST}(\vec{\theta})$, and 
a new scaled deflection, $\vec{\alpha}_\mathrm{MST}$, as
\begin{align}
\label{eq:kappa_tr}
\kappa_\mathrm{MST}(\vec{\theta}) =& \lambda + (1-\lambda) \kappa_\mathrm{model}(\vec{\theta}),\\
\vec{\alpha}_\mathrm{MST}(\vec{\theta}) =& \lambda \vec{\theta} + (1-\lambda) \vec{\alpha}_\mathrm{model}(\vec{\theta})\\
=&\lambda \vec{\theta}+\vec{\alpha}_\mathrm{MST,lens}(\vec{\theta}),
\label{eq:def_tr}
\end{align}
where $\lambda$ is a constant which physically corresponds to the scaled convergence of 
a uniform sheet of mass external to the lens galaxy. In equation (\ref{eq:def_tr}), we decompose
the transformed deflection into two parts; a deflection from the lens, and that from the external convergence. 
We define $\vec{\alpha}_\mathrm{MST,lens} \equiv (1-\lambda)\vec{\alpha}_\mathrm{model}$, 
whose meaning will be explained later in this subsection.
To satisfy the lens equation (\ref{eq:lens}) while leaving the image positions invariant, 
the source position must transform as
\begin{equation}
\vec{\beta}_\mathrm{MST} = (1-\lambda)\vec{\beta}_\mathrm{model},
\end{equation}
which is why this transformation is a part of the family of transformation called the \textit{source-position} transformation.

Considering the following relation among $\kappa$ , $\phi$ and $\psi$, 
\begin{eqnarray}
\phi &=& \frac{1}{2}(\vec{\theta} - \vec{\beta})^2 - \psi,\\
\nabla^2 \psi &=& 2 \kappa, 
\end{eqnarray}
the transformed Fermat potential of the $i$-th image, $\phi_{\mathrm{MST},i}$, becomes
\begin{equation}
\phi_{\mathrm{MST},i}=(1-\lambda)\phi_{\mathrm{model},i}-\frac{\lambda(1-\lambda)}{2}|\vec{\beta}|^2.
\label{eq:fermat_pot}
\end{equation}
Since the source position $\vec{\beta}$ is the same for all the images, the second term
in equation (\ref{eq:fermat_pot}) cancels out if we calculate the difference in the 
Fermat potential between two images $i$ and $j$. Thus, the difference, $\Delta \phi_{i,j}$, transforms as
\begin{equation}
\Delta\phi_{\mathrm{MST},i,j} = (1-\lambda) \Delta\phi_{\mathrm{model},i,j}.
\label{eq:phi_tr}
\end{equation}
As the time delay is directly proportional to the Fermat potential, we find that 
$\Delta t_{i,j}$ is simply increased by a factor of $1-\lambda$ after the MST for fixed distances/cosmology.

If we assume that the physical origin of MST is an effective 
 external convergence due to mass structures along the line of sight, 
 $\kappa_{\rm ext}$, we can identify $\lambda$ with $\kappa_{\rm ext}$.
In the following, we apply the MST to the power-law mass model and 
show that the inferred $D_A(EL)$ remains unaffected by $\kappa_{\rm ext}$.  
We start first with the special case of SIS to gain intuition before considering 
the general power-law profile.


\subsubsection{Singular isothermal sphere}
Here we follow the steps from section \ref{sec:sis}, but with MST
applied to it. From equation (\ref{eq:kappa_tr}), the transformed
density profile of the lens is
\begin{equation}
\rho_\mathrm{SIS,MST} = (1-\kappa_\mathrm{ext})\rho_\mathrm{SIS,model}.
\end{equation}
Note that the original transformation equation (\ref{eq:kappa_tr}) is
written in terms of the convergence, $\kappa$; however, as the
convergence and the density profile are proportional to each other
(equation (\ref{eq:convergence-def})), we transform the density in the same
way as the convergence. To satisfy equation (\ref{eq:rho_sis}), the
velocity dispersion must transform as 
\begin{equation}
\sigma^2_\mathrm{MST} = (1-\kappa_\mathrm{ext})\sigma^2.
\end{equation}
Equation (\ref{eq:sig}) then becomes
\begin{equation}
\sigma^2_\mathrm{MST} = (1-\kappa_\mathrm{ext})\theta_\mathrm{E}\frac{c^2}{4\pi}\frac{D_A(ES)}{D_A(LS)}.
\end{equation}
From equation (\ref{eq:phi_tr}), the time-delay equation (\ref{eq:SISdt}) transforms as
\begin{equation}
\Delta t_{\mathrm{MST},i,j}=(1-\kappa_\mathrm{ext})\frac{1+z_\mathrm{L}}{2c}\frac{D_A(EL) D_A(ES)}{D_A(LS)} (\theta_{j}^2-\theta_{i}^2),
 \end{equation} 
 and by combining the above two equations, we get
\begin{equation}
 \Delta t_{\mathrm{MST},i,j} = \frac{4\pi }{c^3}\sigma^2_\mathrm{MST}(1+z_\mathrm{L})D_A(EL)(\theta_{i}-\theta_{j}),
\label{eq:D_A_SIS_MST}
\end{equation}
in which $\kappa_{\rm ext}$ cancels out. This equation
is identical to equation (\ref{eq:D_A_SIS}), but with the
transformed quantities, $\Delta t_{\mathrm{MST},i,j}$ and
$\sigma^2_\mathrm{MST}$.

The reason is as follows. 
Suppose that we have a lens system which has a velocity dispersion of 
$\sigma^2$ and the time-delay difference of $\Delta t$. 
We then try to model this system by a lens plus an external convergence, 
$\kappa_{\rm ext}$. Then, the modeled $\sigma$ and $\Delta t$ 
would be different from the original ones by a factor of $1-\kappa_{\rm ext}$, 
but the ratio of the two is invariant. As $D_A(EL)$ is proportional
to the ratio of the two, we can measure the same $D_A(EL)$ as before,
regardless of the existence of the external convergence.


\subsubsection{Spherical power-law density profile}
Now we study the effect of MST on the spherical power-law density profile lens
galaxy model, following section \ref{sec:power-law}.
Combining the time-delay transformation with equation (\ref{eq:dt_delta}) yields
\begin{equation}
\begin{aligned}
\Delta t_{\mathrm{MST},i,j} = &(1-\kappa_\mathrm{ext})\Delta t_{\mathrm{model},i,j} \\
= &(1-\kappa_\mathrm{ext})D_A(EL)\frac{(1+z_\mathrm{L})}{2c}\\
&\times\left[ (\hat{\alpha}_{\mathrm{model},i} + \hat{\alpha}_{\mathrm{model},j})\cdot (\vec{\theta}_i - \vec{\theta}_j) -\frac{2}{l}(\vec{\theta_i}\cdot\hat{\alpha}_{\mathrm{model},i} -\vec{\theta_j}\cdot \hat{\alpha}_{\mathrm{model},j} )\right].
\label{eq:dt_D}
\end{aligned} 
\end{equation}
Again, the density normalization of the lens galaxy, $\rho_0$, transforms as
\begin{equation}
\rho_\mathrm{0,MST} = (1-\kappa_\mathrm{ext})\rho_\mathrm{0,model},
\end{equation}
and thus among the total deflection angle $\alpha$, only a $(1-\kappa_\mathrm{ext})$ fraction of it is from the lens,
which is why we denoted this contribution as
$\vec{\alpha}_\mathrm{MST,lens} =
(1-\lambda)\vec{\alpha}_\mathrm{model}$ in equation
(\ref{eq:def_tr}). 
Using this in equation~(\ref{eq:dt_D}) yields
\begin{equation}
\begin{aligned}
\Delta t_{\mathrm{MST},i,j} &=D_A(EL)\frac{(1+z_L)}{2c}\\
\times &\left[ (\hat{\alpha}_{\mathrm{MST,lens},i} +
 \hat{\alpha}_{\mathrm{MST,lens},j})\cdot (\vec{\theta}_{i} -
 \vec{\theta}_{j})
 -\frac{2}{l}(\vec{\theta_{i}}\cdot\hat{\alpha}_{\mathrm{MST,lens},i}
 -\vec{\theta_{j}}\cdot \hat{\alpha}_{\mathrm{MST,lens},j} )\right].
\end{aligned}
\end{equation}
As the measured velocity dispersion of the lens gives the estimate of the lens potential only, the relation between 
the deflection angle from the lens and the velocity dispersion does not change after the MST:
\begin{equation}
\vert\hat{\alpha}_\mathrm{MST,lens}\vert = \frac{4(\gamma'-1)}{c^2}\sigma_r^2(R) F(\gamma')\left(\frac{b}{R}\right)^{-\gamma'+2}.
\end{equation}
Thus, $D_A(EL)$ can be calculated from the original equation (\ref{eq:power-law}) even after the MST. 


This is an important finding. In the previous studies of the time-delay distance to measure the Hubble constant, 
$\kappa_{\rm ext}$ was the main obstacle in measuring $H_0$ precisely \citep{suyu/others:2010}. 
On the other hand, we have shown that $D_A(EL)$ measured from strong
lensing, which combines the time-delay, the image position, and the
velocity dispersion data, does not suffer from the effect of
$\kappa_\mathrm{ext}$.

\section{Error formula and implications for B1608+686 and RXJ1131$-$1231}
\label{sec:implications}
\subsection{Aperture-averaged line of sight velocity dispersion}
\label{sec:ap}
We do not measure the radial component of the velocity dispersion, $\sigma_r^2(r)$.
 Rather, we measure the luminosity-weighted line-of-sight velocity dispersion, $\sigma_p^2(R)$. 
 We relate them using the following equation:
\begin{equation}
\sigma_p^2(R) \equiv I_p(R)\sigma^2_s(R) = 2\int_R^{\infty}\Big(1-\beta_\mathrm{ani}\frac{R^2}{r^2}\Big)\frac{\rho_*(r) \sigma_r^2(r) r dr}{\sqrt{r^2-R^2}}.
\label{eq:sig_s}
\end{equation}
Here, $r$ denotes the three-dimensional radius, while $R$ denotes the projected radius.
We shall use these two different radii notations for the rest of the paper. 
$\beta_\mathrm{ani}$ is the effect of the velocity dispersion anisotropy, which will be
studied in detail in section \ref{sec:anisotropy}. In this section, we set $\beta_\mathrm{ani}=0$.
The other functions are: $I_p(R)$ is the projected stellar distribution function, $\sigma_s(R)$ is the 
projected velocity dispersion and $\sigma_r(r)$ is given by equation (\ref{eq:sig_r}).
For a stellar density profile, $\rho_*$, we consider two profiles that are known to describe 
well the stellar light distributions of galaxies: the Hernquist profile and the Jaffe profile.  
These two different profiles would also allow us to assess the effect of luminosity 
weighting on $\sigma_{\rm p}^2(R)$.

A generalized form of the stellar density distribution, which satisfies $\rho_* \propto r^{-4}$ as $r\rightarrow \infty$, can be expressed as
\begin{equation}
\rho_* = \frac{(3-\gamma_s)I_0}{4\pi}\frac{a}{r^{\gamma_s}(r+a)^{4-\gamma_s}},
\end{equation}
where $0\leq\gamma_s<3$, following ref. \citep{dehnen:1993}. The Hernquist profile 
corresponds to $\gamma_s=1$ \citep{Hernquist:1990}:
\begin{equation}
\rho_*(r) =\frac{I_0 a}{2\pi r(r+a)^3},
\end{equation}
where $I_0$ is a normalization and $a$ is a scale radius determined by $a = (2^{1/(3-\gamma_s)}-1)r_{1/2}$ following ref. \citep{dehnen:1993}, where $r_{1/2}$ is the half-mass radius. Due to the projection effect, the two-dimensional half-light radius $R_\mathrm{eff}$ is related to the three-dimensional half-mass radius $r_{1/2}$ as $r_{1/2} = 1.33 R_\mathrm{eff}$ for the Hernquist profile, thus the scale radius $a =0.551 R_\mathrm{eff}$.
The projected Hernquist distribution is known to provide a
good fit for the stellar distribution of elliptical galaxies that follow
the de Vaucouleurs law,
\begin{equation}
I_p(R) = \frac{I_0}{2\pi a^2(1-s^2)^2}[(2+s^2)X(s)-3],
\end{equation}
where $s \equiv R/a$ is a scaled projected radius, and $X(s)$ is defined as
\begin{eqnarray}
X(s) \equiv \begin{cases}
	\displaystyle\frac{1}{\sqrt{1-s^2}} \operatorname{sech}^{-1} s & \textrm{for} ~0\leq s \leq 1\\
	\displaystyle\frac{1}{\sqrt{s^2-1}} \operatorname{sec}^{-1} s & \textrm{for} ~1\leq s < \infty
	\end{cases}~~~.
\label{eq:x(s)}
\end{eqnarray}
We then examine the case where the stellar density profile has a steeper slope at the center, by using 
the Jaffe model \citep{Jaffe:1983}, which has $\gamma_s=2$. 
We do not consider models with $2<\gamma_s<3$ as they fail to represent the basic physical properties of a galaxy, 
e.g. diverging potential / velocity dispersion at the center. In the Jaffe model, the stellar density profile becomes 
\begin{equation}
\rho_* = \frac{1}{4\pi}\frac{a}{r^{2}(r+a)^{2}},
\end{equation}
and the projected surface brightness distribution, $I_p(R)$, becomes
\begin{equation}
I_p(R) = \frac{I_0}{4a^2s}-\frac{I_0}{2\pi a^2}\frac{1}{s^2-1}\left[(s^2-2)X(s)+1\right],
\end{equation}
where for the Jaffe profile $r_{1/2} = 1.31 R_\mathrm{eff}$ and $a = 1.31 R_\mathrm{eff}$.

We note that both the Hernquist and Jaffe profiles for the stars are not single power-laws, and neither are dark matter distributions such as the Navarro, Frenk and White profile \cite{NFW:1996}. Stars and dark matter have different radial distributions in galaxies with the stars typically dominating over dark matter at the central parts and vice versa at outer parts.  The contributions of stars and dark matter are often comparable around the effective radius. Despite the different radial distributions of stellar and dark matter, the total density profile of stars and dark matter is remarkably well described by a power-law within a few effective radius, as previous lensing and/or dynamical studies have shown (e.g.,\cite{koopmans/etal:2009, barnabe/etal:2011, suyu/others:2013, cappellari/etal:2015}). Therefore, our use of the Hernquist/Jaffe profiles for the luminosity weighting to scale the velocity dispersion measured near the effective radius to the image positions is consistent with the use of a power-law for the total density profile. For a total density profile that is nearly isothermal, there is an inconsistency in the slope between the total density profile and the Hernquist profile at the center ($r \lesssim 0.1′′$). However, the contribution of the enclosed mass from this central region to the total enclosed mass within either the Einstein radius or effective radius (approximately where we have lensing/dynamical measurements) is insignificant.  Thus, the central slope inconsistency between the stellar and the total density has negligible impact on our lensing and dynamical analysis.

Ideally, we wish to measure the line-of-sight velocity dispersion profile as a function of projected radii.
In practice, however, most of the observations do not allow us to spatially resolve
the galaxy; rather they allow us to measure the luminosity-weighted, 
aperture-averaged velocity dispersion inside an aperture of a fixed size 
\cite{cappellari/others:2006}. 
We calculate the luminosity-weighted aperture-averaged projected
velocity dispersion, $\langle\sigma_p^2\rangle_{\rm ap}$, as follows:
\begin{equation}
\begin{aligned}
\langle\sigma_p^2\rangle_{\rm ap} \equiv~&\frac{\int_{\rm ap}
 I_p\sigma_s^2 R ~dR ~ d\theta}{\int_{\rm ap} I_p R ~dR ~d\theta}.
\end{aligned}
\label{eq:sig_p}
\end{equation}
 \subsection{Analytic formula}
In this section, we relate the statistical uncertainty in $D_A$ to those
of the observables, i.e., $\Delta t$, $\sigma^2_p$, and $\gamma'$. (The
effect of an anisotropic velocity dispersion will be discussed in detail in
section~\ref{sec:anisotropy}.) Assuming that these observables are
independently measured, we write the total uncertainty in $D_A(EL)$,
hereafter $S_{D_A}$, as
\begin{equation}
\begin{aligned}
S_{D_A} =& \sqrt{\left(\frac{\partial D_A}{\partial \Delta t_{i,j}}\right)^2 S_{\Delta t}^2 + \left(\frac{\partial D_A}{\partial \sigma_p^2}\right)^2 S_{\sigma_p^2}^2 + \left(\frac{\partial D_A}{\partial \gamma'}\right)^2 S_{\gamma'}^2} \\
=&D_A\sqrt{\left(\frac{1}{\Delta t_{i,j}}\right)^2 S_{\Delta t}^2 + \left(\frac{1}{\sigma_p^2}\right)^2 S_{\sigma_p^2}^2+\frac{1}{D_A^2}\left(\frac{\partial D_A}{\partial \gamma'}\right)^2 S_{\gamma'}^2},
\label{eq:Sda}
\end{aligned}
\end{equation} 
where $S_x$ is the measurement uncertainty in the variable $x$. 
Since image positions, $\theta_{i,j}$, are precisely measured,  we do
not include their uncertainties in this formula. 
In the following sections \ref{sec:b1608} and \ref{sec:rxj1131}, we
shall apply this formula to two lens systems, B1608+656 and
RXJ1131$-$1231, respectively.

\subsection{B1608+656}
\label{sec:b1608}
\begin{figure}[t]
\begin{center}
\includegraphics[height=6cm]{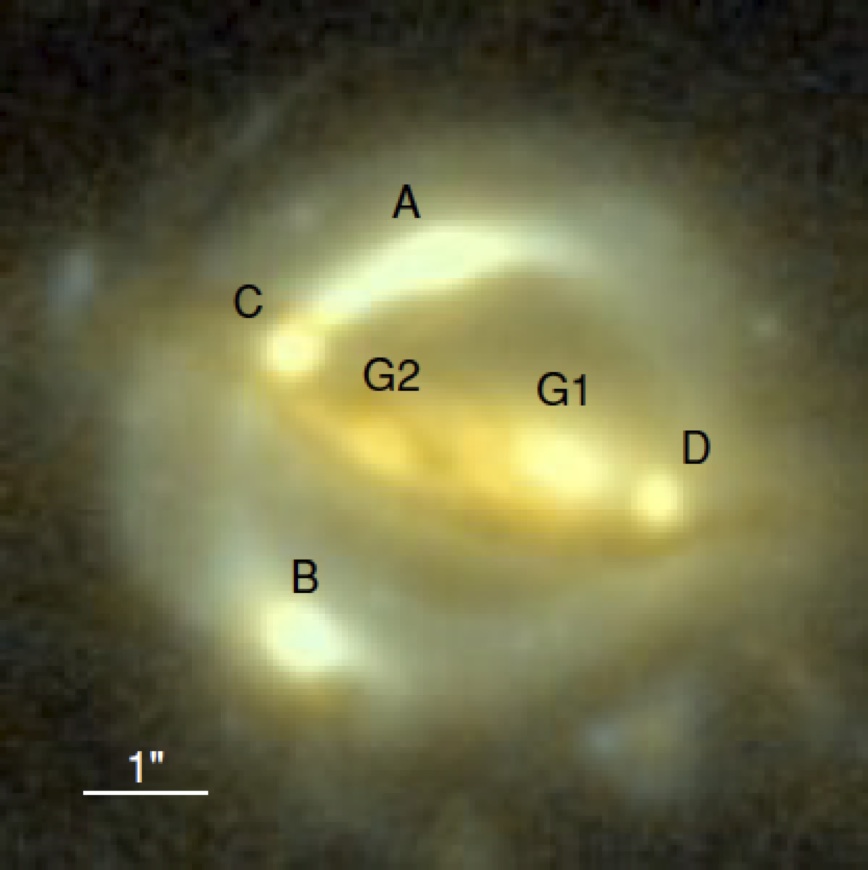}
\caption{Image of B1608+656, adopted from figure 1 of \cite{suyu/others:2010}.}
\label{fig:B1608+656}
\end{center}
\end{figure}

Figure \ref{fig:B1608+656} shows the image configuration of B1608+656
\citep{suyu/others:2009}. The information on image configuration is
important as our formula applies only to a circularly symmetric
case. Thus, the only image pairs we can use in this paper are the ones
that are on the opposite sides of the lens center. More
thorough analysis using all the data will be presented elsewhere (Suyu
et al., in preparation).
The data of B1608+656 are mostly from ref. \cite{suyu/others:2010}, but the image positions are
 calculated from the data given in ref. \cite{koopmans/etal:2003}, the time delays are from ref. \cite{fassnacht/others:2002}, 
 and the redshifts are from refs. \cite{fassnacht/others:1996,myers/others:1995}. 
 For this system, the origin of the coordinates is set at the image A. The data are summarized as :
\begin{equation}
\begin{aligned}
z_\mathrm{L} =& 0.6304\\
z_\mathrm{s}=&1.394\\
\vec{\theta}_\mathrm{A} =& ( 0.0'',0.0 '')\\
\vec{\theta}_\mathrm{B} =& ( -0.7380'',-1.9612 '')\\
\vec{\theta}_\mathrm{C}=& (-0.7446'',-0.4537 '')\\
\vec{\theta}_\mathrm{D} =&(1.1284 '',-1.2565 '')\\
R_\mathrm{eff} =& 0.58''\\
\gamma' =& 2.08 \pm 0.03\\
\langle\sigma_p^2\rangle_\mathrm{ap}^{1/2} =& 260 \pm 15 ~ \mathrm{km/s}\\
\Delta t_\mathrm{AB} =& 31.5 ^{+2.0}_{-1.0}~ \mathrm{days}\\
\Delta t_\mathrm{CB} =& 36.0 ^{+1.5}_{-1.5}~ \mathrm{days}\\
\Delta t_\mathrm{DB} =& 77.0 ^{+2.0}_{-1.0}~ \mathrm{days}\\
\Delta t_\mathrm{CD} = & \Delta t_\mathrm{CB} - \Delta t_\mathrm{DB} = -41.0 ^{+2.5}_{-1.8}~ \mathrm{days}.
\end{aligned}
\end{equation}
We use the CD pair. Also, as we write $D_A(EL)$ in terms of $\sigma_r(r)$ 
(e.g. \ref{eq:power-law}), we normalize the radial velocity dispersion profile, 
$\sigma_r(r)$, using $\langle\sigma^2_p(R)\rangle_{\rm ap}$ 
given by the observation and using equations (\ref{eq:sig_s}) and (\ref{eq:sig_p}).
With these values, we find $D_A(EL) = 1485.7~\mathrm{Mpc}$.
For comparison, $D_A(EL)$ from the best-fit WMAP 7-year parameters is $D_A(EL)
= 1423.3~\mathrm{Mpc}$. We now use equation (\ref{eq:Sda}) to
compute $S_{D_A}$:
\begin{equation}
S_{D_A}= D_A\sqrt{3.72\times10^{-3}+
1.33\times10^{-2}+2.36\times10^{-3}},
\label{eq:SDA}
\end{equation}
where from the first term, each number indicates the fractional uncertainty in $D_A$ contributed by the time-delay measurement $\Delta t_{i,j}$, the line-of-sight velocity dispersion measurement $\sigma_p^2$, and the density profile index 
$\gamma'$. (Note that
$S_{\sigma_p^2}/\sigma_p^2=2S_{\sigma_p}/\sigma_p$.) 
With this value, the total uncertainty, including all the terms in equation
(\ref{eq:SDA}), is $S_{D_A}=0.14D_A$, i.e., 14\% uncertainty. 
The dominant contribution comes from the uncertainty in $\sigma_p$,
which gives $S_{D_A}=0.12D_A$.

\subsection{RXJ1131$-$1231}
\label{sec:rxj1131}
In this section we repeat the same analysis as above, but with another well-studied strong lensing time-delay system, RXJ1131$-$1231, using data from refs. \cite{tewes/others:2013, Sluse/others:2003} for the time delays and the redshifts, respectively, and from ref. \cite{suyu/others:2012} for the other quantities.
\begin{figure}[t]
\begin{center}
\includegraphics[height=6cm]{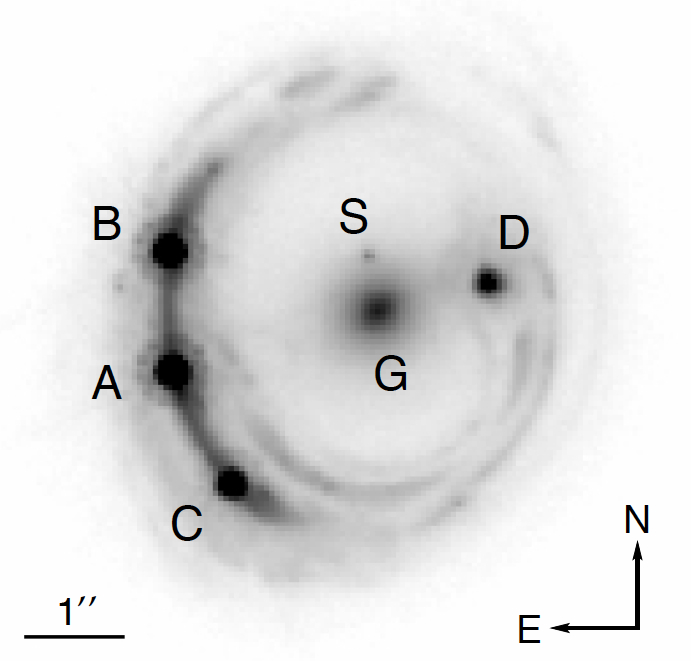}
\caption{Image of RXJ1131$-$1231, adopted from figure 1 of \cite{suyu/others:2012}.}
\label{fig:RXJ1131-1231}
\end{center}
\end{figure}
The data for this system are
\begin{equation}
\begin{aligned}
z_\mathrm{L} =& 0.295\\
z_\mathrm{s}=&0.658\\
\vec{\theta}_\mathrm{G} =&(4.411'',4.011'')\\
\vec{\theta}_\mathrm{A}=& (2.383'',3.412'')\\
\vec{\theta}_\mathrm{D} =& (5.494'',4.288'') \\
R_\mathrm{eff} =&1.85''\\
\gamma' =& 1.95 ^{+0.05}_{-0.04}\\
\langle\sigma_p^2\rangle_\mathrm{ap}^{1/2} =& 323 \pm 20~ \mathrm{km/s}\\
\Delta t_\mathrm{AB} =& 0.7 \pm 1.4 ~\mathrm{days}\\
\Delta t_\mathrm{DB} =& 91.4 \pm 1.5 ~\mathrm{days}\\
\Delta t_\mathrm{AD} =& \Delta t_\mathrm{AB} - \Delta t_\mathrm{DB} =
 -90.7 \pm 2.1~ \mathrm{days}.
\end{aligned}
\end{equation}
We use the AD pair. Using these values, we find $D_A(EL) = 813.33
~\mathrm{Mpc}$, and $D_A(EL)$ from the best-fit WMAP 7-year parameters is
$D_A(EL) = 876.5 ~\mathrm{Mpc}$. The total uncertainty in $D_A(EL)$ is 
\begin{equation}
S_{D_A}= D_A\sqrt{5.36\times10^{-4} +1.53\times10^{-2}+1.46\times10^{-3}}= 0.13 D_A.
\end{equation}
The velocity dispersion alone gives $S_{D_A}=0.12D_A$.

Therefore,  we expect the {\it existing} data on these
systems to yield $D_A(EL)$ with $13-14\%$ precision per object, assuming the
isotropic velocity dispersion. In the next section, we shall study the
effect of the largest source of systematic uncertainty in our method: an
anisotropic velocity dispersion, and how to reduce its effect in the
estimation of $D_A(EL)$.

\section{Anisotropic velocity dispersion}
\label{sec:anisotropy}
The anisotropic stellar motion changes the relation between the
potential and the observed line-of-sight velocity
dispersion. As our method crucially relies upon knowing the
potential depth, we must take into account the anisotropic velocity
dispersion of stars. We do this by following ref.~\cite{suyu/others:2010}, 
which uses  spherical Jeans modeling to relate the observed line-of-sight velocity dispersion to the 
mass distribution. We then study the effect of
anisotropy on the aperture-averaged value of the velocity dispersion
(section~\ref{sec:jeans}) as well as on the velocity dispersion measured at the
so-called ``sweet spot'' (section~\ref{sec:sweet_spot}). Finally, we
use Monte Carlo simulations to compute the effect of anisotropy on the
uncertainty in $D_A(EL)$ (section~\ref{sec:mc}).

\subsection{Spherical Jeans equation}
\label{sec:jeans}
We solve the spherical Jeans equation for a given mass distribution (i.e., a power-law density profile)
 to obtain the three-dimensional radial velocity dispersion $\sigma_r$,
\begin{equation}
\frac{1}{\rho_*}\frac{d(\sigma_r^2 \rho_*)}{dr} + 2 \beta_\mathrm{ani}
 \frac{\sigma_r^2}{r} = -\frac{GM(\le r)}{r^2}.
\label{eq:jeans}
\end{equation}
Here, the anisotropy function, $\beta_\mathrm{ani}(r)$, is chosen as the
Osipkov-Merritt anisotropy \cite{osipkov:1979,merritt:1985}, 
\begin{equation}
\beta_\mathrm{ani}(r) \equiv \frac{r^2}{r_\mathrm{ani}^2 + r^2} = 1-\frac{\sigma_T^2(r)}{\sigma_r^2(r)},
\label{eq:o-m_ani}
\end{equation}
where $\sigma_T(r)$ and $\sigma_r(r)$ are the velocity dispersions in
the tangential  and radial directions, respectively. Although the
anisotropy is parametrized by a single variable, $r_{\rm ani}$, under
this specific model, we can model almost any velocity structures  by
linearly superimposing the solutions \cite{merritt:1985}. We then
calculate $\sigma^2_p(R)$ from $\sigma_r^2(r)$ using equation
\ref{eq:sig_s}, and $\langle\sigma_p^2\rangle_{\rm ap}$ using equation~\ref{eq:sig_p}.

\begin{figure}[t]
\begin{center}
\includegraphics[height=7cm]{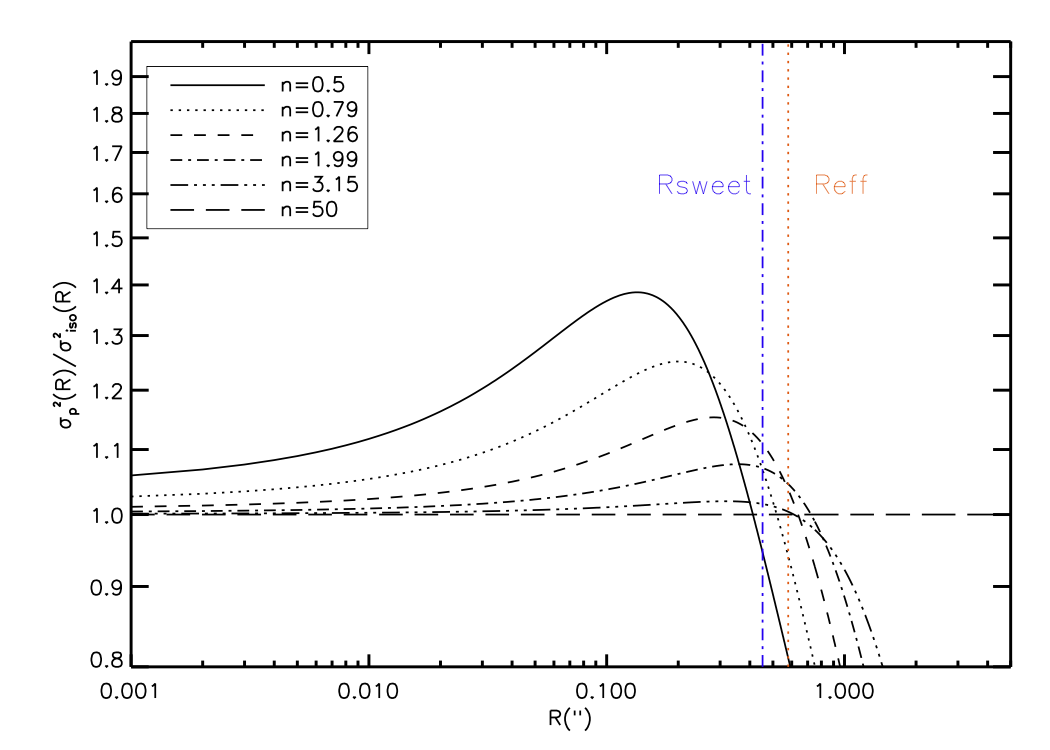}
\caption{Ratio of $\sigma_p^2(R)$ to $\sigma_\mathrm{iso}^2(R)$, as a
 function of the projected radius $R$, and $n \equiv r_{\rm ani}/R_{\rm eff}$. 
 The former is observable,
 while the latter is related more directly to $GM/R$. Two vertical lines
 show the effective radius ($R_{\rm eff}$) and the sweet-spot radius
 ($R_{\rm sweet}$) defined in section~\ref{sec:sweet_spot}.
}
\label{fig:vel_rat}
\end{center}
\end{figure}
To quantitatively demonstrate the behavior of the anisotropic velocity dispersion, 
we again use the observations of B1608+656 introduced in section \ref{sec:b1608} 
for the analysis in this and the following sections.

In figure \ref{fig:vel_rat}, we show the ratio of $\sigma_p^2(R)$ to the
isotropic velocity dispersion, $\sigma_\mathrm{iso}^2(R)$, with $a=0.551
R_{\mathrm{eff}}$ and $R_{\mathrm{eff}} = 0.58''$ for the Hernquist profile. The isotropic
velocity dispersion is a solution to the Jeans equation (\ref{eq:jeans}) with no
anisotropy, $\beta_{\rm ani}\equiv 0$; thus, it is related more directly
to $GM/R$. We have one free parameter, $n$, which parametrizes the
anisotropic radius as
\begin{equation}
 r_{\mathrm{ani}} \equiv n R_{\mathrm{eff}}. 
 \label{eq:rani}
\end{equation}
For a given mass distribution of the lens,
 $\sigma_p^2(R)$ depends on $n$. We vary $n$ from 0.5 to 50 in
 logarithmic spacing. We find $\sigma_p^2(R)/\sigma_{\rm
 iso}^2(R)\approx 1$ to within 10\%  at $R=R_{\rm eff}$, except for the
 highly anisotropic case of $n=0.5$ when the stellar distribution follows Hernquist profile.

\begin{figure}[t]
\begin{minipage}[b]{0.45\linewidth}
\centering
\includegraphics[height=5.5cm]{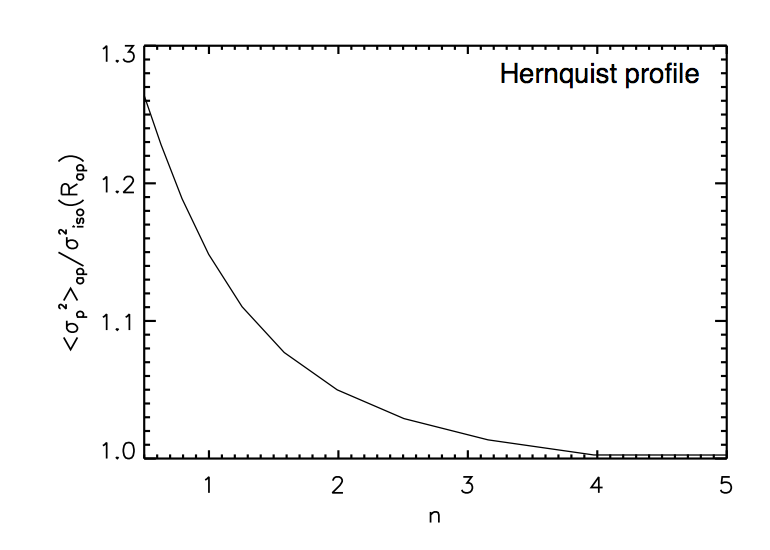}
\end{minipage}
\hspace{0.5cm}
\begin{minipage}[b]{0.45\linewidth}
\centering
\includegraphics[height=5.5cm]{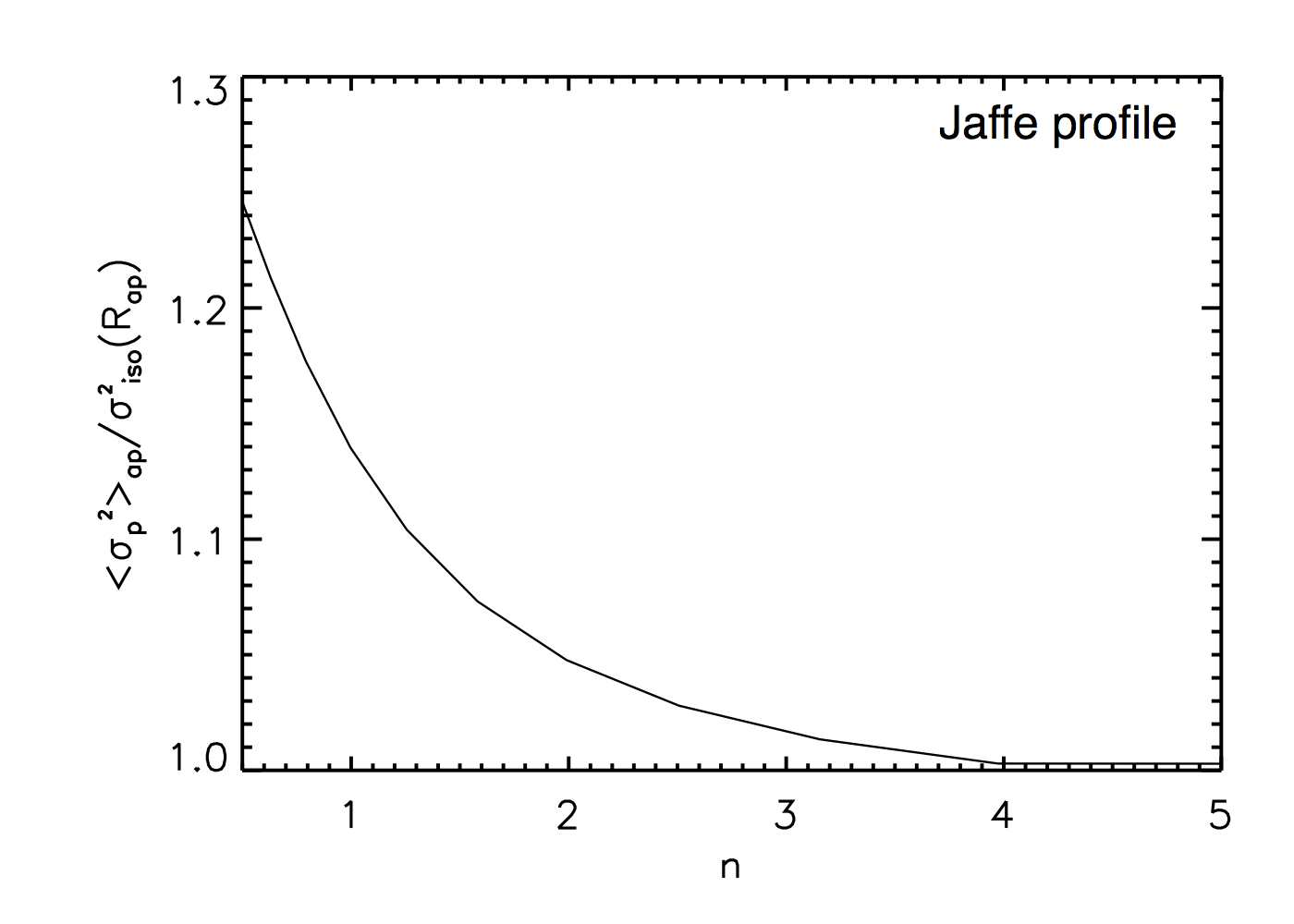}
\end{minipage}
\caption{Ratio of $\langle\sigma_p^2\rangle_{\rm ap}$ to
 $\sigma_\mathrm{iso}^2(R_{\rm ap})$, as a function of $n$. For the stellar
 density distribution, we use the 
 Hernquist and Jaffe profiles in the left and right panels, respectively.
 The size of the aperture is fixed at $R_\mathrm{ap}=0.42''$. The upper
 limit of $n$, 5, is chosen since the velocity dispersion does not differ 
 much from the isotropic case beyond $n$ of 5, while the lower limit, 
 0.5, is determined by observations (e.g. \cite{kronawitter/others:2000})
 and radial instability arguments (e.g. \cite{merritt/aguilar:1985, stiavelli/sparke:1991}).
 The unknown anisotropy dominates the uncertainty on $\langle\sigma_{\rm p}^2\rangle_{\rm ap}$.}
\label{fig:norm_B}
\end{figure}

In figure \ref{fig:norm_B}, we show the ratio of
$\langle\sigma_p^2\rangle_{\rm ap}$ to $\sigma_\mathrm{iso}^2(R_{\rm ap})$ as a
function of $n$, where $R_{\rm ap}$ is fixed to 0.42". In the left panel, this ratio reaches 26\% for $n=0.5$, 
and decreases as $n$ increases when the stellar distribution follows the Hernquist profile.
 In the right panel, we show the same ratio for the Jaffe stellar distribution, with the ratio reaching 24\% for $n=0.5$.  
 Overall the difference in $\langle \sigma_p^2\rangle_\mathrm{ap}$ between Hernquist and Jaffe
 distributions is small compared to the impact of the anisotropy.  Therefore, for the remainder of the paper,
 we consider only the Hernquist distribution as a conservative model, where the dominant uncertainty on $\langle\sigma_p^2\rangle_\mathrm{ap}$ 
 is due to the unknown anisotropy. 

Since the inferred $D_A(EL)$ is proportional to the
 inverse of the isotropic velocity dispersion,
having a large variation in the inferred isotropic velocity dispersion
can cause a large uncertainty in $D_A(EL)$.
Unfortunately, anisotropy is not directly observable, unless we have 
a three-dimensional velocity dispersion measurement. Clearly, a better
approach is needed.

\subsection{Sweet-spot method}
\label{sec:sweet_spot}

It has been pointed out that, when the observations of the surface
brightness profile and the velocity dispersion profile are available,
one can find the so-called \textit{sweet spot}, $R_\mathrm{sweet}$, at
which the effect of the anisotropic velocity dispersion on the mass
determination is minimized \cite{Churazov/others:2010}. Also see
\cite{wolf/others:2010, walker/others:2009}. The
Osipkov-Merritt anisotropy model has an isotropic core and a radial
envelope. However, as we observe the projected velocity dispersion, there are two components that
play roles in the estimation of the observed velocity dispersion. 
The anisotropy changes the ratio between tangential and radial components of 
the velocity dispersion at a given radius, 
while the projection changes the magnitude of contributions from radial and tangential components. 
Quantitatively, at a fixed radius of observation $R$, $\sigma_p(R)$ has contributions from 
infinitely many shells with radii $r=R/\mathrm{cos}\,x$, where $x=[0,\pi/2]$. 
At each radius $r$ we can decompose the contributions to the projected velocity dispersion 
into tangential and radial components as 
$\sigma_T(r)\,\mathrm{cos}\,x$ and $\sigma_r(r)\,\mathrm{sin}\,x$, respectively.
Due to the weighting by the trigonometric functions, at small $x$, 
contributions from the tangential component is bigger than that from the 
radial component, and vice versa at large $x$.
Now, let us assume that the total velocity dispersion, $\sigma^2(r) = \sigma_T^2(r)+\sigma_r^2(r)$, 
is the same for the isotropic and anisotropic model for a given galaxy mass, as it is proportional to 
the total kinetic energy. Then, when $r=R/\mathrm{cos}\,x$ is large, the tangential component 
is suppressed while the radial component is enhanced compare to the isotropic case, 
due to the anisotropy (since $\sigma_T(r)$ becomes small for large $r$ in equation (\ref{eq:o-m_ani})). 
As a result, in comparison to the isotropic case where $\sigma_T(r) = \sigma_r(r)$, 
anisotropic velocity dispersion shows $\sigma_p(R)>\sigma_{\rm iso}(R)$ at small $R$, and 
$\sigma_p(R)<\sigma_{\rm iso}(R)$ at large $R$.
Thus, if we observe an anisotropic system, there exists a projected radius $R$ at which the 
transition from one to the other occurs, as we increase $R$ 
from the center of a galaxy to the outskirt of it. 
This transition radius is the sweet spot.

 While the analytical derivation of
$R_\mathrm{sweet}$ has been done assuming a constant
$\beta_\mathrm{ani}$, the further study \cite{Lyskova/others:2012} shows that the method works for
systems with a non-constant $\beta_\mathrm{ani}$ as well. 
The sweet spot can be determined from the brightness profile of a massive elliptical galaxy 
\cite{Lyskova/others:2012}. It is close to the projected radius at which
 $R$ satisfies $d\ln I(R)/d\ln R = -2$. 
For a Hernquist surface brightness profile, 
we find $R_\mathrm{sweet} \approx 0.78R_\mathrm{eff}$.
It is also shown in \citep{Churazov/others:2010} that while the Sersic index changes from 1 to 12, 
$R_\mathrm{sweet}$ varies only about $0.3 R_\mathrm{eff}$, thus the sweet-spot radius is fairly 
insensitive to the luminosity profile. In figure
\ref{fig:vel_rat}, the sweet-spot radius is shown as the left
vertical line. We find that the difference between projected velocity
dispersions with various anisotropy parameters is minimum around
$R=0.78R_\mathrm{eff}=0.45''$ with the data of B1608+656. It particularly reduces the
effect of a highly anisotropic case with $n=0.5$, compared to  using
$\sigma^2_p(R)$ at the effective radius or the aperture-averaged
$\sigma_p^2$. The uncertainty in the mass of massive ellipticals
estimated from the sweet-spot method is claimed to be 5-7 per
cent. Therefore, the best approach is to use
spatially-resolved spectroscopic data of lens galaxies to obtain the
velocity dispersion at the sweet spot.

\subsection{Monte Carlo simulation}
\label{sec:mc}
We use Monte Carlo simulations to study how much the
velocity anisotropy inflates the uncertainty in $D_A(EL)$, and how well
we can mitigate it by using the sweet-spot method.

For two time-delay systems B1608+656 and RXJ1131$-$1231, we generate 11
discrete radial profiles of anisotropic velocity dispersion by solving
the Jeans equation, with logarithmically spaced $n=[0.5,50]$. The
effective radius and the density profile index $\gamma'$ are fixed at the
best-fit values given in sections~\ref{sec:b1608} and
\ref{sec:rxj1131}. We randomly choose one profile from the set of
different anisotropy parameters to create a mock galaxy. We
then compute $\sigma_p^2$ from each mock galaxy in three ways: the
aperture-averaged value $\langle\sigma_p^2\rangle_{\rm ap}$ with the
aperture size of $0.42''$ for both systems, $\sigma_p^2(R)$ with
$R=R_{\rm eff}$, and $\sigma_p^2(R)$ with $R=R_{\rm sweet}$. As the
uncertainty in $D_A(EL)$ is dominated by that of $\sigma_p^2$, we add a
Gaussian random noise to $\sigma_p^2$ with variance of 
$S_{\sigma^2_p}^2 = 2 S_{\sigma_p}^2(S_{\sigma_p}^2+2{\sigma}^2_p)$. We
then compute $D_A(EL)$ from these simulated data with the best-fit
values of the time-delay data and image positions given in
sections~\ref{sec:b1608} and \ref{sec:rxj1131}. (We do not add noise to
time delays or image positions.)  While our simulated
galaxies have anisotropic velocity dispersions, we use the isotropic
velocity dispersion model to calculate $D_A(EL)$. In this way we can
quantify the effect of our ignoring anisotropic velocity dispersion by
marginalizing over it.

\begin{figure}[t]
\centering
\includegraphics[height=7cm]{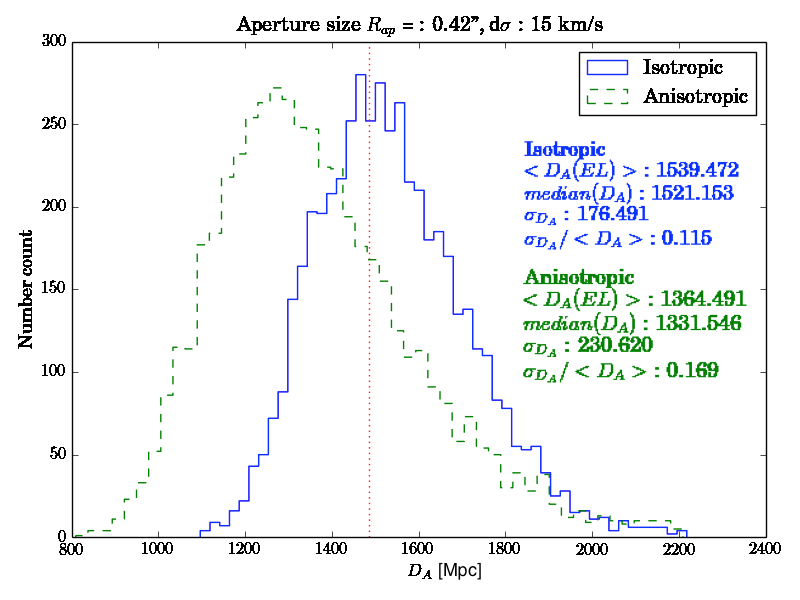}
\caption{Simulated distribution of $D_A(EL)$ to B1608+656. The solid
 and dashed histograms show the distributions with the isotropic and
 anisotropic simulations, interpreted by the isotropic model.
We use the aperture averaged velocity dispersion,
 $\langle\sigma_p^2\rangle_{\rm ap}$, with the aperture size of
 $0.42''$. The standard deviation of the velocity dispersion used in
 simulations is $15 ~\mathrm{km/s}$. The fractional uncertainty in $D_\mathrm{A}$ is
 11.5\% in the case of isotropic velocity dispersion model, while in the case of anisotropic velocity dispersion model
 it is 16.9\%.}
\label{fig:1608_d_a_rap}
\end{figure}

\begin{figure}[t]
\centering
\includegraphics[height=7cm]{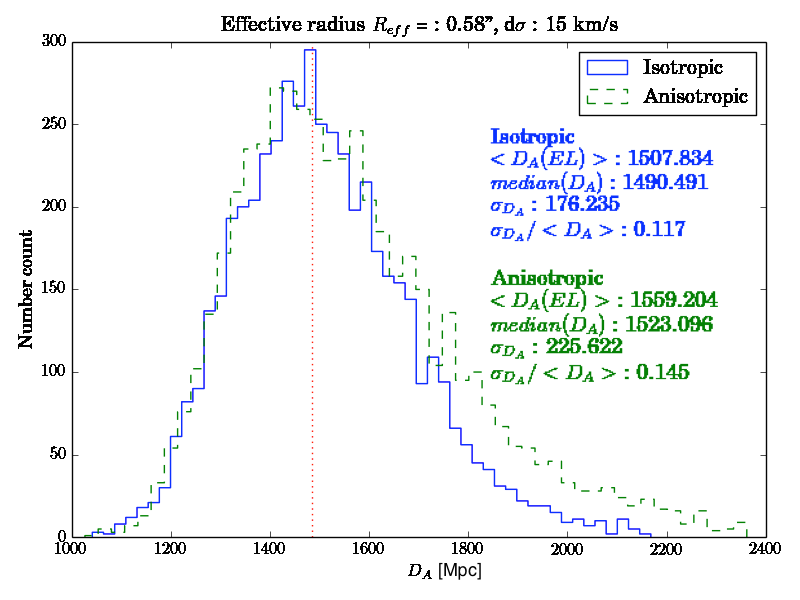}
\caption{Same as figure~\ref{fig:1608_d_a_rap}, but with
 $\sigma_p^2(R_{\rm eff})$ and $R_\mathrm{eff} = 0.58''$. The fractional uncertainty in $D_\mathrm{A}$ is
 11.7\% in the case of isotropic velocity dispersion model, while in the case of anisotropic velocity dispersion model
 it is 14.5\%.}
\label{fig:1608_d_a_ref}
\end{figure}

\begin{figure}[t]
\centering
\includegraphics[height=7cm]{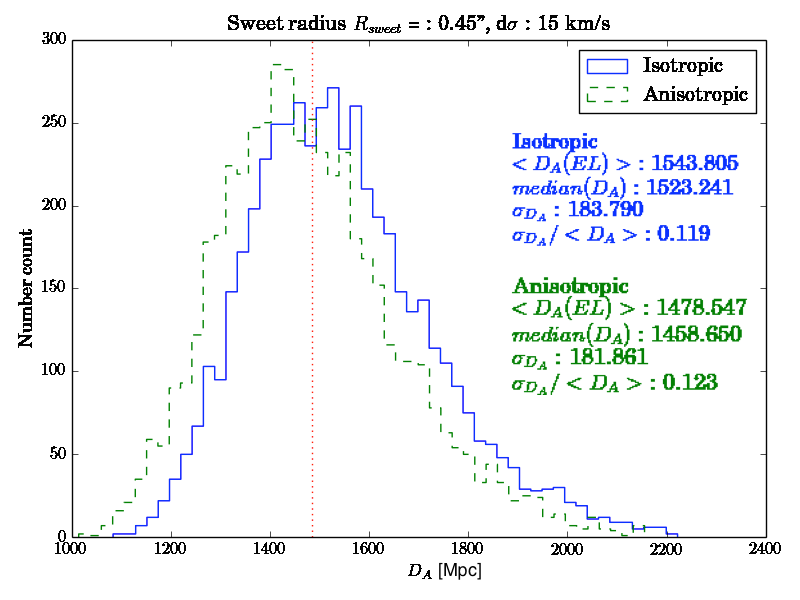}
\caption{Same as figure~\ref{fig:1608_d_a_rap}, but with
 $\sigma_p^2(R_{\rm sweet})$ and $R_\mathrm{sweet} = 0.45''$. The fractional uncertainty in $D_\mathrm{A}$ is
 11.9\% in the case of isotropic velocity dispersion model, while in the case of anisotropic velocity dispersion model
 it is 12.3\%.}
\label{fig:1608_d_a_rsw}
\end{figure}

Figures \ref{fig:1608_d_a_rap}, \ref{fig:1608_d_a_ref}, and
\ref{fig:1608_d_a_rsw} show the distributions of $D_A(EL)$ obtained from
mock B1608+656 realizations using $\langle\sigma_p^2\rangle_{\rm ap}$,
$\sigma_p^2(R_{\rm eff})$, and $\sigma_p^2(R_{\rm sweet})$,
respectively. The solid and dashed histograms in each panel show the
realizations with the isotropic and anisotropic velocity dispersions,
respectively. The former realizations are used to check validity of our
simulations, as well as to make a direct assessment of the effect of
anisotropy. The vertical dotted lines show $D_A(EL)=1485.7$~Mpc that we
obtained in section~\ref{sec:b1608}.


We summarize the results from the analysis on B1608+656 and RXJ1131$-$1231
in tables \ref{tb:1608} and \ref{tb:1131}, respectively. The
uncertainties in $D_A(EL)$ from isotropic simulations (interpreted by
the isotropic model) agree with the analytical estimates given in
sections~\ref{sec:b1608} and \ref{sec:rxj1131}. 
On the other hand, those from anisotropic simulations (again
interpreted by the isotropic model) show significantly larger uncertainties when
$\langle\sigma_p^2\rangle_{\rm ap}$ or $\sigma_p^2(R_{\rm eff})$ is
used. Fortunately, using $\sigma_p^2(R_{\rm sweet})$ eliminates most of
the inflation of the uncertainty due to velocity anisotropy. \\
Figure \ref{fig:1608_d_a_rsw} shows that the peak is shifted in the anisotropic case in comparison to
the isotropic case, while in figure \ref{fig:1608_d_a_ref} the peak remains at the same position. 
This is due to the marginalization of the anisotropy. 
In figure \ref{fig:vel_rat}, we choose 6 different $n$ values that are spaced logarithmically, 
and choose two radii ($R_{\rm eff}$ and $R_{\rm sweet}$) to calculate the
$D_A$ distributions for both the isotropic and anisotropic cases. 
At $R_{\rm sweet}$, the scatter between the curves is smaller compare to
that at $R_{\rm eff}$; however, at $R_{\rm sweet}$, the curves are also shifted toward higher
velocity dispersions compared to the isotropic case. As a result, the whole distribution of $D_A$ is shifted
toward lower values. On the other hand, at $R_{\rm eff}$, while
the scatter is larger, there is no systematic change in $\sigma_p^2$ relative to $\sigma_{\rm iso}$ 
(i.e. among 6 values of $n$, two give $\sigma_p^2$ 
larger than the $\sigma_{\rm iso}^2$, two give smaller, and the other two give $\sigma_p^2$
almost identical to the $\sigma_{\rm iso}^2$ value). As a result, the peak position does not change,
while we get an extended tail towards higher $D_A$ value. This does not mean that using 
$R_{\rm sweet}$ gives a biased $D_A$, as we cannot assume that the velocity dispersion structure
is isotropic. Also, as the width of the distribution is much bigger than the shift of the peak, 
at the moment the effect of this shift is negligible.
As the distribution of $D_A$ depends on the choice of the anisotropy model 
as well as on the range/selection of $n$, we study another anisotropy parameterization 
to see the robustness of the results against the choice of parameterization in the next section.
\begin{table}[t]
\centering
\caption{Expected fractional uncertainty in $D_A(EL)$ to B1608+656}
\begin{tabular}{ccc}
  \toprule[1.5pt]
  \head{} & \head{Isotropic} & \head{Anisotropic}\\
  \midrule
 $R_\mathrm{ap}$ &11.5\%& 16.9\%\\
  $R_\mathrm{eff}$ & 11.7\%& 14.5\%\\
  $R_\mathrm{sweet}$ &11.9\% & 12.3\%\\
  \bottomrule[1.5pt]
\end{tabular}
\label{tb:1608}
\end{table}
\begin{table}[t]
\centering
\caption{Expected fractional uncertainty in $D_A(EL)$ to RXJ1131$-$1231}
\begin{tabular}{ccc}
  \toprule[1.5pt]
  \head{} & \head{Isotropic} & \head{Anisotropic}\\
  \midrule
  $R_\mathrm{ap}$ & 12.5\% & 15.1\% \\
  $R_\mathrm{eff}$ & 12.4\% & 14.8\% \\
  $R_\mathrm{sweet}$ &12.5\%& 12.6\% \\
  \bottomrule[1.5pt]
\end{tabular}
 \label{tb:1131}
\end{table}

\subsection{Agnello et al. (2014) parameterization}
To show that the sweet spot is not a unique characteristic of Osipkov-Merritt anisotropy,
we repeat the same analysis using a different spatially-varying anisotropy parameter, $\beta_\mathrm{ani}(r)$,
from ref.\citep{agnello/etal:2014} :
\begin{equation}
\beta_\mathrm{ani}(r) = \frac{\beta_\mathrm{in} r^2 + \beta_\mathrm{out} r_\mathrm a^2}{r^2 + r_\mathrm a^2}.
\label{eq:adri_param}
\end{equation}
Two additional parameters, $\beta_\mathrm{in}$ and 
$\beta_\mathrm{out}$, are added to the Osipkov-Merritt anisotropy. We follow ref.\citep{agnello/etal:2014} and
adopt flat priors on $\beta_\mathrm{in} = [-0.6,0.6]$ and $\beta_\mathrm{out}=[-0.6,0.6]$,
while the anisotropic radius, $r_\mathrm a$, is scaled in the same way as in the Osipkov-Merritt model (equation \ref{eq:rani}).

\begin{figure}[t]
\begin{center}
\includegraphics[height=8cm]{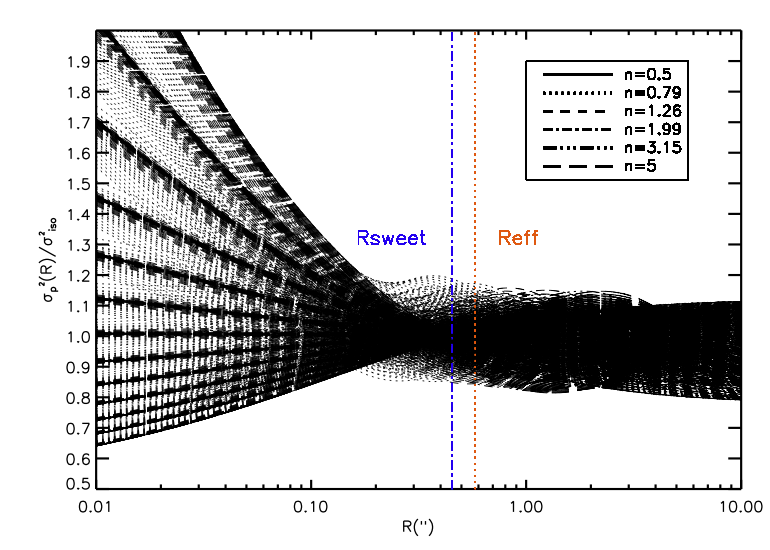}
\caption{Same as figure \ref{fig:vel_rat}, but with another anisotropic velocity dispersion parametrization given by
equation \ref{eq:adri_param}. The range of the two new parameters, $\beta_{\rm in}$ and $\beta_{\rm out}$, is $[-0.6,0.6]$ for both parameters, with steps of $\delta \beta_{\rm in}=0.1$ and  $\delta \beta_{\rm out}=0.1$. } 
\label{fig:adri_sig_profile}
\end{center}
\end{figure}

\begin{figure}[t]
\begin{center}
\includegraphics[height=8cm]{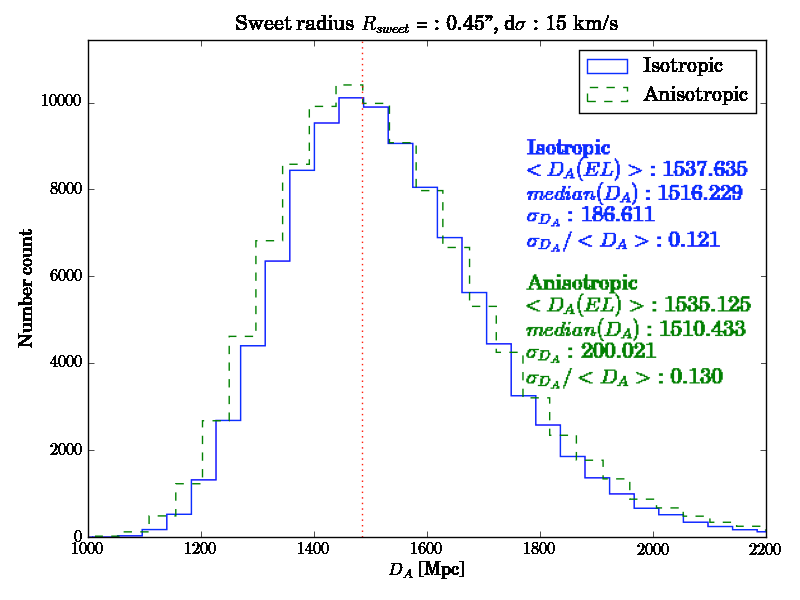}
\caption{Same as figure \ref{fig:1608_d_a_rsw}, but with the anisotropic velocity dispersion parametrization
given by equation \ref{eq:adri_param}. Note that we use more realizations here in comparison to the 
previous analysis, as the parameter combination is 169 times as much as the one from 
Osipkov-Merritt parameterization, due to two additional parameters.  As a consequence, 
the result for the isotropic case is slightly different from figure \ref{fig:1608_d_a_rsw}. The fractional uncertainty in $D_\mathrm{A}$ is
 12.1\% in the case of isotropic velocity dispersion model, while in the case of anisotropic velocity dispersion model
 it is 13.0\%.} 
\label{fig:adri_da_rsw}
\end{center}
\end{figure}

The resulting velocity dispersion profiles are shown in figure \ref{fig:adri_sig_profile}. 
Near the sweet spot, the fractional uncertainty in the velocity dispersion becomes as small as $15\%$. 
Also we note that the deviation from the isotropic velocity dispersion is not skewed at the sweet spot, which keeps
the peak of the posterior distribution of $D_A$ at the same place as for the isotropic
dispersion model. The posterior distribution calculated at the sweet-spot radius
is shown in figure \ref{fig:adri_da_rsw}. We find that the uncertainty on angular diameter distance using
this parametrization is about $13\%$ for B1608+656, and $14\%$ for RXJ1131$-$1231, comparable to those in section \ref{sec:mc}.

\section{Conclusion}
\label{sec:discussion}

\begin{figure}[t]
\begin{center}
\includegraphics[height=8cm]{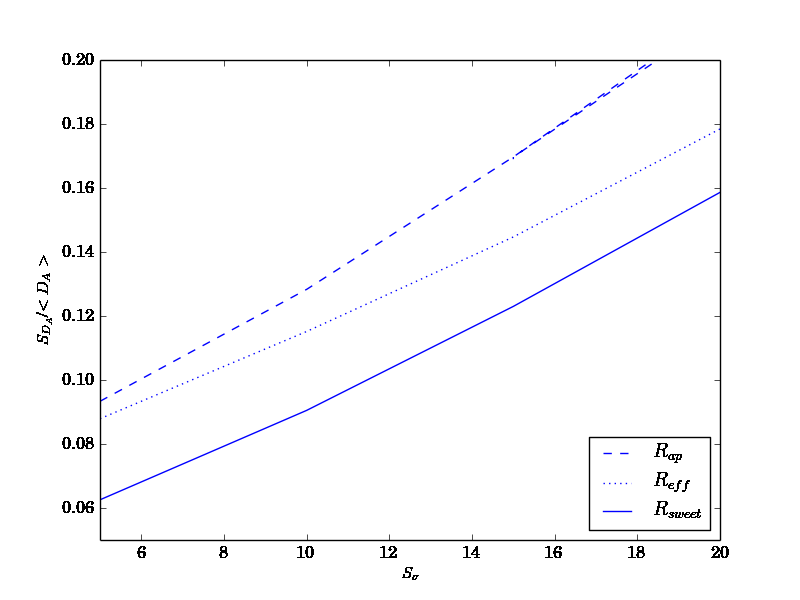}
\caption{Expected fractional uncertainty in $D_A(EL)$ to B1608+656 as a
 function of the  uncertainty in $\sigma$ in units of
 km/s. The dashed,  dotted, and solid lines are for $\sigma=\langle
 \sigma_p^2\rangle^{1/2}_{\rm ap}$, $\sigma_p(R_{\rm eff})$, and
 $\sigma_p(R_{\rm sweet})$, respectively.} 
\label{fig:dsig_dda}
\end{center}
\end{figure}

In this paper, we have shown that we can determine $D_A(EL)$ to strong
lens systems with time delays. The underlying physics is simple; thus,
this method offers a robust determination of $D_A(EL)$ to individual
systems. The key advantage of this method is that the external
convergence does not affect the distance determination. The uncertainty
in the inferred $D_A(EL)$ is dominated by that in the velocity dispersion
and its anisotropy. The effect of anisotropy can be minimized by
measuring the velocity dispersion at the sweet-spot radius.

The {\it existing} data on B1608+656 and RXJ1131$-$1231 should yield
$D_A(EL)$ with 17\% and 15\% precision, respectively. If we use the
velocity dispersions at the sweet-spot radii, the precision improves to about
13\%. In figure~\ref{fig:dsig_dda}, we show the expected fractional uncertainty
in $D_A(EL)$ to B1608+656 as a function of the uncertainty in the
velocity dispersions, $\sigma$. The $\sigma$ at the sweet-spot radius
measured with $260\pm 7$~km/s corresponds to $\sigma^2$ measured with
5\% precision. This yields $D_A(EL)$ with 7\% precision, after
marginalizing over velocity anisotropy. We show the robustness of our results using two different parameterizations
of $\beta_\mathrm{ani}(R)$, but a further study may be needed for more general cases.

This paper describes the basic idea and presents an estimate of what we
can do with the existing data. Since we assumed spherical density
profiles, our analysis is not precise enough to yield the best
determinations of $D_A(EL)$ to B1608+656 and RXJ1131$-$1231. The method
presented in this paper has been implemented in the full analysis
pipeline used by refs.~\cite{suyu/others:2010,suyu/others:2012}, and the
results will be reported in a future publication (Suyu et al., in preparation).

\acknowledgments
IJ would like to thank Karl Gebhardt for his support and
encouragement, and Eugene Churazov, Adriano Agnello, Matthew Auger, Simona Vegetti,
Stefan Hilbert, Chiara Spiniello, Natalia Lyskova, and Edward Robinson
for useful discussions. EK would like to thank Phil Marshall and Masamune Oguri 
for useful discussions, and Stefan Hilbert for his clear lecture on
strong lenses. SHS would like to thank Adriano Agnello, Matthew Auger and Matteo Barnab\`e
for helpful discussions. Funding for this work has been provided in part by 
Texas Cosmology Center (TCC). TCC is supported by the College of Natural 
Sciences and the Department of Astronomy at the University of Texas at Austin 
and the McDonald Observatory.
\bibliography{151009}
\appendix
\section{Deflection angle of an arbitrary power-law density profile}
\label{sec:def_angle}
We derive the expression for a deflection angle near a galaxy with the
density profile following a power-law with arbitrary density profile index.
When the density profile is given as equation (\ref{eq:density}), the mass contained within a radius $r$ is
\begin{equation}
M(r) = \int_{0}^r4 \pi r^2 \rho_0 r_0^{\gamma'} r^{-\gamma'}dr = \frac{4\pi \rho_0 r_0^{\gamma'}}{-\gamma' + 3}r^{-\gamma'+3},
\end{equation}
which yields an acceleration given by
\begin{equation} 
\vec{g}(\vec{r})=-\frac{4\pi G \rho_0 r_0^{\gamma'}}{3-\gamma'}r^{-\gamma'}\vec{r},
\end{equation}
on the test mass located at the radius $r$. According to the post-Newtonian approximation 
in General Relativity, the rate of change of the direction of the velocity vector of the test mass, $\vec{u}$, is given as
\begin{equation}
c^2 \frac{d\vec{u}}{dt} = -2\vec{u} \times (\vec{u} \times \vec{g}).
\end{equation}
We define a new parameter $\alpha$ to be the angle by which the light is deflected 
as it passes near the lens galaxy. In the cases we consider, the deflection angle will be small. 
Thus, we can choose coordinates such that the path of the light is roughly along the x-axis, 
and the line connecting the center of the lens galaxy to the 
point of the closest approach is along the y axis. Again, because the deflection angle 
is small, we use the thin lens approximation, namely, light is bent sharply at the 
closest approach to the lens. Thus, the separation from the center of lens to the light path, 
$r$, becomes $r^2 =b^2 + x^2 $, and, more importantly, 
$\vec{u} \times (\vec{u} \times \vec{r}) = - c^2 \vec{b}$. 

We define the deflection angle at the lens plane, $\hat{\alpha}$, as the total change in the photon propagation direction, and the magnitude of the deflection angle as $\alpha$. Then,
\begin{equation}
\hat{\alpha} \equiv \frac{1}{c}\int d\vec{u} = -\alpha \, \frac{\vec{r}}{r},
\end{equation}
 where the minus sign indicates that the deflection happens toward the lens center. Then $\alpha$ becomes
\begin{equation}
\begin{aligned}
\alpha &= \frac{8\pi G \rho_0 r_0^{\gamma'}} {c^2(3-\gamma')} \int _{-\infty}^{\infty}b r^{-\gamma'}dx\\
&= \frac{8\pi G \rho_0 r_0^{\gamma'}} {c^2(3-\gamma')}b \int _{-\infty}^{\infty}\frac{dx}{(x^2 + b^2)^{\gamma'/2}}\\
&= \frac{8\pi G \rho_0 r_0^{\gamma'} b^{2- \gamma'}}{c^2(3-\gamma')}\frac{ \sqrt{ \pi} ~\Gamma[ \frac{1}{2}(-1+\gamma')]}{\Gamma(\frac{\gamma'}{2})},
\end{aligned}
\end{equation}
assuming $\gamma' >1$.
\end{document}